# Terahertz for Radar applications and Wireless Communication

1st S. Latreche
Université de Bejaia, Faculté de
Technologie, Laboratoire de Génie
Electrique, 06000 Bejaia, Algeria
sofiane.latreche@univ-bejaia.dz

2nd H. Bellahsene
Université de Bejaia, Faculté de
Technologie, Laboratoire de Génie
Electrique, 06000 Bejaia, Algeria
hocine.bellahsene@univ-bejaia.dz

3rd A. Taleb-Ahmed
Valenciennes University
IEMN DOAE Laboratory
abdelmalik.taleb-ahmed@uphf.fr

*Abstract*—Technological advancements in the design of electronic and optical materials have opened up the possibility of utilizing the latest available Radio Frequency spectrum - the Terahertz (THz) band. This band holds great promise for next-generation wireless systems, which are poised to seamlessly integrate a wide array of data-intensive and time-sensitive applications. In this article, we delve into the Terahertz band, providing insights into its properties and showcasing examples of its applications. We begin by exploring the specific characteristics of wireless communications and radar systems operating in the THz band. Subsequently, we analyze various effects and parameters unique to each of these applications.so we scrutinize the application of Terahertz (THz) wireless and radar systems, delving into the modeling of various facets of radio frequency propagation within this domain. The interpretation of our findings will be presented at the conclusion of this study.

## I. INTRODUCTION

Mobile networks, whether in the radio part or the core part, are evolving very fast. Starting with a generation dedicated to telephony until the arrival of applications that are unimaginable for most people. Those applications need high speed and low latency. In order to meet these requirements, the terahertz band is considered as the only key technology until now. Terahertz (THz) waves, or sub-millimeter/far-infrared waves, refer to electromagnetic radiation with inside the frequency interval from 0.1 to ten THz. They occupy a large part of the electromagnetic spectrum [1].

The THz frequency guarantees extensive bandwidth, which theoretically reaches as much as numerous THz, resulting in a potential capacity with inside the order of Terabits per second [2]. Hence, the bandwidth provided is one order of magnitude above millimeter-wave (mmW) systems. THz signals additionally permit better hyperlink directionality and provide lower eavesdropping probabilities while as compared to their millimeter counterparts [3]. Analysis of the THz band suggests that those frequencies additionally ownfixed of benefits in comparison to optical frequencies. In uplink THz waves are a serious candidate for radio transmission. They allow non-line-of-sight (NLoS) propagation [4] and act as good substitutes below inconvenient weather situations such as fog, dust and turbulence [5]. In addition, the THz frequency band isn't always impacted with the aid of using ambient noise arising from optical sources, neither is it related to any health restrictions or safety limits [6]. Table 1 presents a comparison among the THz frequency band and different current technologies.

In particular, the harsh propagation environment features a high path loss inversely proportional to the square of the wavelength, and thus to the size of a single antenna element, and, in addition, a high molecular absorption in certain frequency bands [7]. the smallest wavelength at terahertz allows many antenna elements to be packed in a small surface, thus enabling Ultra-massive Multiple Input, Multiple Output (UM-MIMO) techniques [7]. several studies have focused on increasing the communication range in macro scenarios [8], and on signal generation and modulation. Directional antennas are used to mitigate the increased pathloss, as they can focus the power in narrow beams, which increase the link budget, and to enhance the security of wireless links [9].

In this paper, we first provide an in-depth exploration of Terahertz (THz) frequencies, followed by a comprehensive analysis of their suitability for mobile networks. Subsequently, we delve into the examination of their applicability in detection radar systems, considering various scenarios such as free space and fog conditions. The study includes the calculation of Signal-to-Noise Ratio (SNR) for different MMWave and Terahertz frequencies to assess the capacity of THz frequencies and the required antenna configurations for THz utilization. Finally, we present the findings and engage in a discussion of the results. A conclusive summary will be provided at the conclusion of our research.

## II. THE THz FREQUENCIES

THz-band communications are predicted to play a pivotal position in the imminent 6th generation (6G) of wireless mobile communications, allowing ultra-high-bandwidth communication paradigms. To this end, many research groups have attracted significant financial funding to conduct THz research. The objective is to boost the standardization efforts and to reduce the time allocated to these procedures to a minimum [10].

As the terahertz (THz) band is the last untapped band in the radio frequency (RF) spectrum, technologies from the neighboring microwave and optical bands have been deployed enabling the design of future THz communication techniques

| Technology | mmWave | THz Band | Infrared | Visible Light Communication (VLC) | Ultra-Violet |
|---|---|---|---|---|---|
| Frequency Range | 30 GHz - 300 GHz | 100 GHz - 10 THz | 10 THz - 430 THz | 430 THz - 790 THz | 790 THz - 30 PHz |
| Range | Short range | Short/Medium range | Short/Long range | Short range | Short range |
| Power Consumption | Medium | Medium | Relatively low | Relatively low | Expected to be low |
| Network Topology | Point to Multi-point | Point to Multi-point | Point to Point | Point to Point | Point to Multi-point |
| Noise Source | Thermal noise | Thermal noise | Sun/Ambient Light | Sun/Ambient Light | Sun/Ambient Light |
| Weather Conditions | Robust | Robust | Sensitive | - | Sensitive |
| Security | Medium | High | High | High | To be determined |

TABLE I: Comparison among different wireless communication technologies

[11]. Figure 1 represents the frequency range from 0.1 to 10 THz called the THz band.

The evolution in transceivers allows the use of THz bands in wireless communications. This evolution is mainly electronic and photonic. While photonic technologies have an advantage in terms of data rate and electronic platforms are superior in their ability to generate higher power. [12]

Currently, different technologies are considered to be used in THz transceivers. Like Silicon Germanium (SiGe) technology, compound semiconductor technologies such as Gallium Nitride (GaN) and Indium Phosphide (InP), Photonic devices, such as Quantum Cascade Lasers (QCLs) and bolometric detectors [2]. Plasmonic solutions in particular graphene [13]. Recently, graphene based solutions have emerged as strong candidates that enable communications at the THz band. The particular electric properties of graphene, which include high electron mobility, electrical tunability, and configurability, permit supporting high-frequency signals [14].

The THz band guarantees to help higher user densities, higher reliability, much less latency, greater energy efficiency, higher positioning accuracy, better spectrum utilization, and elevated adaptability to propagation scenarios [12]. During the beyond decade, diverse THz technology had been present process fast improvement and feature enabled us to use THz sensing and imaging for chemical, biological, biomedical and different interdisciplinary study [15]. In current years, THz technology have confirmed to be promising methods for defense and security applications [1]. Also paving the manner for applications with inside the THz band, ranging from indoor wireless communications to automobile and drone communications, to device-to-device communications and nano-communications. In addition, THz signals have the capability for use in lots of non-communications based applications, such as spectroscopy of small biomolecules and quality control of pharmaceuticals [13](Figure 1).

### III. THz and Wireless Communications

THz-band communications will be important to the future 6G and beyond [16]. In particular, the large available bandwidths withinside the order of tens as much as a hundred gigahertz (GHz) and extremely short wavelengths provide enormous potentials to relieve the spectrum scarcity and break the capacity limitation of 5G networks, thereby permitting emerging applications that call for an explosive quantity of data [17]. In addition an ultra-broadband antennas are needed for THz Band communication [18]. Moreover, very large antenna arrays will be necessary to overcome the very high path loss in the THz Band [2]. Below we present simulations of SNR and SNR coverage probability for different antenna configurations and frequencies.

Figure 2 compares the Signal to Noise ratio at totally different distances (5, 50 and 150 m) for THz and mmWave links.

To analyze the impact of the upper carrier frequency on the propagation loss. The SNR is an indicator of the quality of transmission of information. It's given by the ratio between the received signal power and also the noise power, without the beamforming gain, Ptx = 0.5 W, and noise figure F = 10 dB. The bandwidth is 400 megahertz for mmWaves (the maximum bandwidth per carrier of 3GPP NR), and 50 gigahertz for terahertz (compliant with IEEE 802.15.3d).

The SNR is computed mistreatment the 3GPP model in an urban canyon situation within the frequency range considered for 3GPP NR [19], and [20] for THz links in the 300 GHz-1 THz spectrum. We observe the difference between the THz and mmWve carrier. For example, the SNR gap for a distance of 5 m, is 39 dB between a carrier at 41 GHz and at 700 GHz. This difference can be compensated for by increasing the number of antenna elements in each node.

In the Figure 3, we consider simulations during which macro base stations are randomly deployed outdoors according to a Poisson point process, with the coverage probability computed as the probability of getting an Signal to Noise ratio higher than a threshold of zero decibel between a test user, and at least one base station. The path loss and beamforming gain are modeled as in Figure 2.

This figure illustrates the differences in deployment density that will be expected for mmWave and THz systems. It can be seen that with 16 antenna elements at the BS and 4 at the mobile device, the 40 GHz network has a coverage probability greater than 0.95 with 80 BS/km². The same antenna configuration does not guarantee an adequate performance in the terahertz band, since a similar coverage probability requires much more than 1000 base stations/km² at 0.40 THz. But with 1024 antenna elements for base stations and 256 for mobile devices, it is possible to achieve this coverage probability with 100 BS/km² at 0.40 THz and 1100 BS/km² at 2 THz. So as we saw in the previous section, antenna arrays with a larger number of antenna elements (small high density cells) will be needed to remedy the higher path loss in the terahertz frequency spectrum.

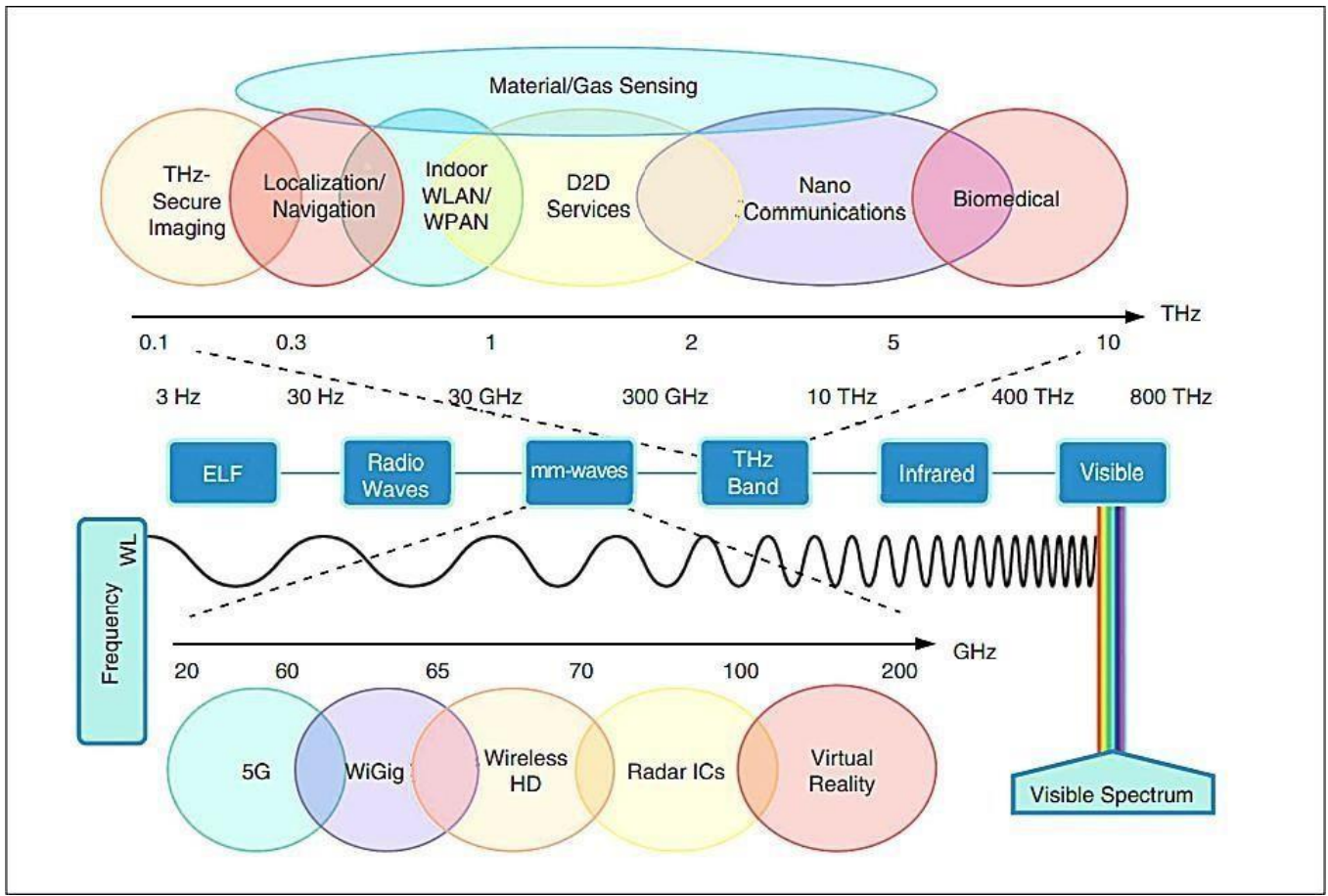

Fig. 1: Evolution of mobile wireless systems.

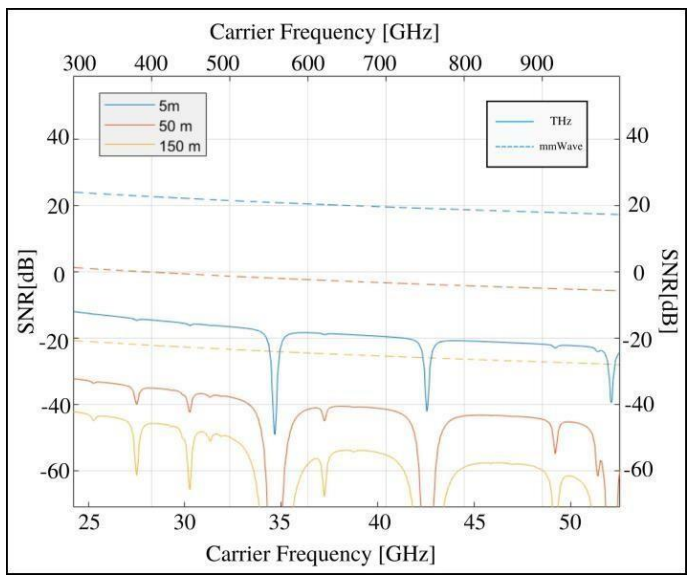

Fig. 2: Millimeter waves and terahertz links

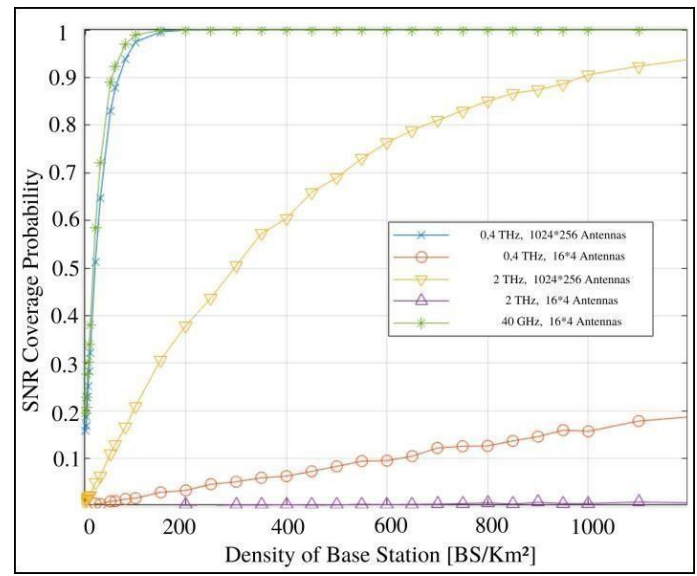

Fig. 3: Probability of getting an SNR higher than 0 dB for various deployment configurations

## IV. THZ AND RADAR SIGNALS

Radar is associate acronym for radio detection and ranging. It uses electromagnetic wave propagation and communication technology to perform distance, angle, and/or rate measurements on targets of interest [21]. To properly assess the performance of radar and wireless communication systems, it's essential to grasp the propagation environment [21]. In this section, we display how to model numerous Radio Frequency propagation effects. These include free space path loss, atmospheric attenuation because of rain, and fog. The examples are based on the ITU-R P-series recommendations of the International Telecommunication Union. The ITU-R is the Radiocommunication Section, and the P-Series deals with radio wave propagation.

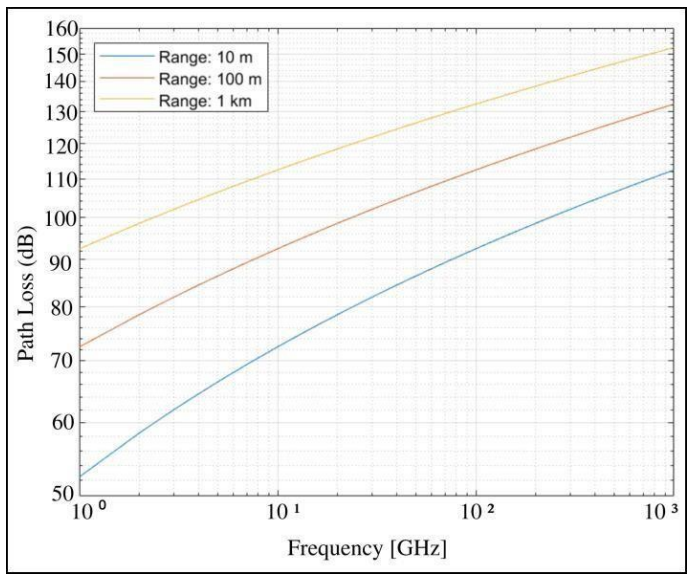

Fig. 4: Free Space Path Loss

In free space, radio frequency signals propagate at the speed of light in all directions. At a sufficient distance, the radiating source resembles a point in space and the wavefront forms a sphere whose radius is equal to R. The path loss in free space is calculated as a function of the propagation distance and the frequency. It is also given by [22]:

$$L_{fs} = 20 * \log_{10}(\frac{4\pi R}{\lambda})\ dB \dots\dots\dots\dots(1)$$

Propagation loss is often expressed in dB. Propagation loss is often expressed in dB. In equation (1), '$\lambda$' is the wavelength and 'R' is the propagation distance.

in our simulation the path loss is determined by factors such as the distance between the transmitter and receiver and the frequency of the signal. The 'fspl' function utilizes these parameters to calculate the path loss for each combination of range and frequency. The resulting path losses are then plotted on a logarithmic scale, which allows for better visualization of the data.

Figure 4 shows the free space path loss (dB) for frequencies between 1 GHz and 1 THz, for different ranges. It can be seen that in the 50 m range, the propagation loss increases from 52 dB to 112 dB proportionally with the [1, 1000] GHz frequency range. From 92.4 dB to 153 dB for the 1 km range. So it can be seen that propagation loss increases with range and frequency.

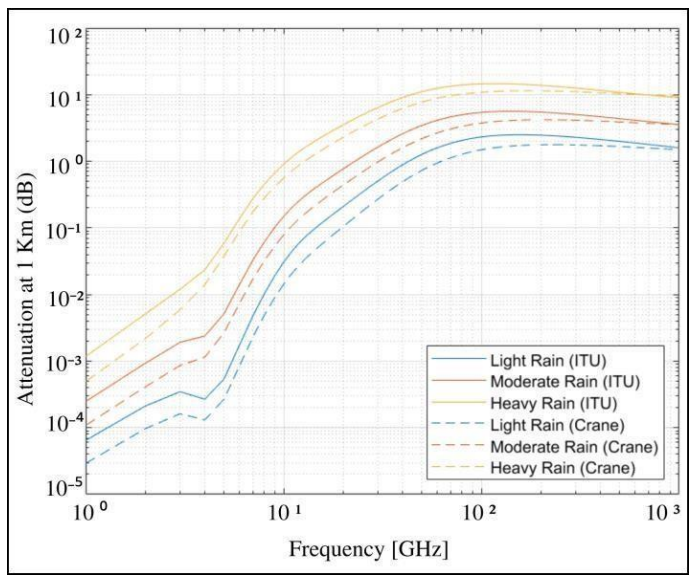

Fig. 5: Rain Attenuation for Horizontal Polarization at 1 km

In reality, signals don't continually travel during a vacuum, therefore free space path loss describes solely a part of the signal attenuation. Signals interact with particles in the air and lose energy along the propagation path. The loss varies with various factors corresponding to pressure, temperature, and water density.

Rain is a significant limiting issue for radar systems, particularly once operational on top of five GHz. Within the ITU model, rain is characterized by the rain rate (in mm/h). According to [23], the rain rate can range from lower than 0.25 mm/h for Very light rain to quite 50 mm/h for extreme rainfall. In addition, because of the shape of the raindrop and its relative size to the wavelength of the RF signal, the propagation loss due to rain is additionally a function of the signal polarization. In general, horizontal polarization represents the worst case of rain propagation loss.

our simulation focuses on calculating and visualizing the Loss Due to Rain. To do this, the code specifies a range and defines an array called 'rainrate' that contains three different rain rates. The simulation then enters a loop where it calculates the rain attenuation for each rain rate using two different functions: 'rainpl' and 'cranerainpl'. These functions take into account the specified parameters, such as range, frequency, rain rate, elevation, and polarization, to calculate the rain attenuation. After calculating the rain attenuations, the resulting values are plotted on a logarithmic scale. This allows for a clear visualization of the impact of rain on signal attenuation.

In Figure 5, we plot losses due to rain calculated with 2 models for a distance of one km. The polarization is assumed to be horizontal, therefore the tilt angle is 0, and therefore the signal propagates parallel to the ground. These two models are the ITU model and Crane. they're valid between 1 GHz and 1 THz. It can be seen that in the case of light rain, the propagation attenuation increases from $3 \cdot 10^{-5}$ dB and $6.5 \cdot 10^{-5}$ dB to 1.6 dB, for the Crane and ITU model, respectively. This increase is proportional with the Frequency range of [1, 1000] GHz. From $5 \cdot 10^{-4}$ dB and $1.2 \cdot 10^{-3}$ dB up to 9.7 dB for the heavy rain case. The losses calculated with the ITU model are generally larger than those calculated with the Crane model at this propagation distance. For longer propagation distances and higher frequencies, the Crane model can produce a larger attenuation value than the ITU model.

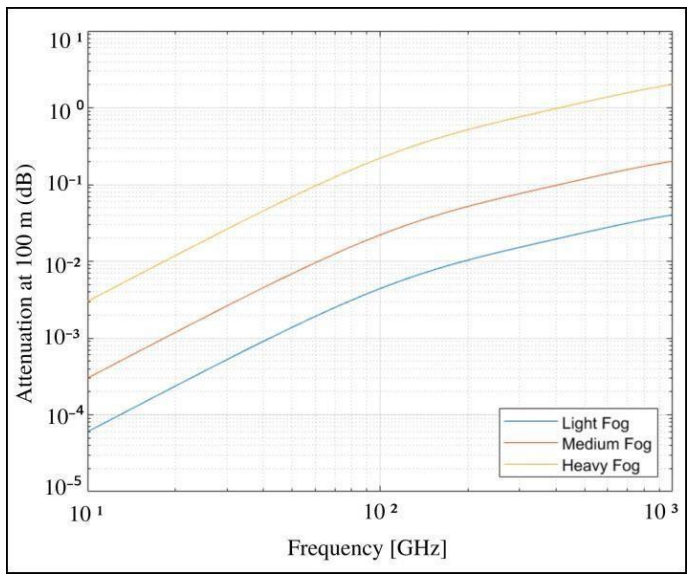

Fig. 6: Fog Attenuation at 100 m

Fog and cloud are shaped with water droplets too, though much smaller compared to rain drops. the dimensions of fog droplets is usually below 0.01 cm. Fog is usually characterised by the liquid water density. A medium fog with a visibility of roughly 300 meters, has a liquid water density of 0.05 $g/m^3$. For heavy fog wherever the visibility drops to 50 meters, the liquid water density is concerning 0.5 $g/m^3$. The atmosphere temperature (in Celsius) is additionally present within the ITU model for propagation loss due to fog and cloud. In general fog isn't present once it's raining [24].

our simulation focuses on calculating and visualizing the Loss Due to Fog. Firstly, a new frequency range is defined, spanning from 10 GHz to 1000 GHz. The variable 'T' represents the temperature in degrees Celsius, which is an important factor in determining fog attenuation. Additionally, the array contains three different liquid water densities, representing varying levels of fog intensity. The code then enters a loop where it calculates the fog attenuation for each water density using the 'fogpl' function. This function takes into account parameters such as range, frequency, temperature, and water density to calculate the fog attenuation. Once the fog attenuations are calculated, the resulting values are plotted on a logarithmic scale. This logarithmic representation helps visualize the impact of fog on signal attenuation more effectively.

In Figure 6, we plot the losses due to fog calculated with the ITU model for a distance of 100 m. This ITU model for fog loss is valid between 10 GHz and 1 THz. It can be seen that in the case of light fog, the propagation loss increases from $6 \cdot 10^{-5}$ dB to 0.04 dB in proportion to the frequency range of [10, 1000] GHz. From 0.003 dB to 2 dB in the case of strong fog. It can therefore be seen that the propagation attenuation increases with increasing fog and frequency.

## V. CONCLUSION

Given the potential of THz communication systems to provide high data rates over short distances, they are currently regarded as the next frontier of research in wireless communications.

This paper explores the utilization of terahertz frequencies in various contexts. Firstly, it investigates their application in mobile networks through simulations, wherein the Signal-to-Noise Ratio (SNR) is calculated across different frequency ranges, distances, and antenna deployments. Secondly, it examines their use in radar systems, modeling several aspects of radio frequency propagation, including free space path loss and atmospheric attenuation due to rain and fog.

In conclusion, it is determined that to effectively use the THz frequency range in mobile networks, antenna arrays with a larger number of antenna elements (UM-MIMO) will be required to compensate for the higher propagation loss within this spectrum. Additionally, THz frequencies are found to exhibit robustness against adverse weather conditions.

ACKNOWLEDGEMENT

The authors would like to thank the director of the Cerist Be´ja¨ıa section Pr. A. Belaid and his team for their welcome in their infrastructure, their technical support and their critical discussions.